\documentclass[aps,pre,showpacs,raggedbottom,nobalancelastpage,amssymb,twocolumn,groupedaddress]{revtex4}
\usepackage{graphicx}
\usepackage{amsmath}
\usepackage{amsfonts}
\usepackage{amssymb}
\usepackage{epsfig,libertine}
\usepackage[libertine]{newtxmath}
\usepackage{color}
\usepackage{dsfont, hyperref, xcolor}
\usepackage{comment}
\usepackage{enumerate}

\newcommand{\numberset}{\mathbb}

\newcommand{\R}{\numberset{R}}

\newcommand{\Tr}[1]{\text{Tr}\left\{#1\right\}}

\newcommand{\bra}[1]{\langle#1\vert}
\newcommand{\ket}[1]{\vert#1\rangle}
\newcommand\braket[2]{\langle#1|#2\rangle}

\begin{document}

\title{What is the most general class of quasiprobabilities of work?}

\author{Gianluca~Francica}
\email{gianluca.francica@gmail.com}
\address{Dipartimento di Fisica e Astronomia ``G. Galilei'', Universit\`{a} degli Studi di Padova, via Marzolo 8, 35131 Padova, Italy}

\date{\today}

\begin{abstract}
How to give a statistical description of thermodynamics in quantum systems is an open fundamental question. Concerning the work, the presence of initial quantum coherence in the energy basis can give rise to a quasiprobability of work, which can take negative values. Our aim is to identify the most general quasiprobability of work satisfying some fundamental conditions. By doing so, we introduce a general notion of quasiprobability in analogy to the Gleason's theorem. Then, we use these quasiprobabilities to define the quasiprobability of work, and finally we discuss the contextuality of the protocol.
\end{abstract}

\maketitle

\section{Introduction}
Thermodynamics can be derived by using a statistical description of nature. However, how to give such description in quantum systems still remains an open problem of fundamental importance. In particular, several attempts have been made to describe the work statistics, after the two-projective-measurement scheme has been originally proposed~\cite{talkner07}. For instance, among these, work distributions have been defined in terms of a work operator~\cite{deffner16}, gaussian measurements~\cite{talkner16}, full-counting statistics~\cite{solinas15}, weak values~\cite{allahverdyan14} and consistent histories~\cite{miller17}.
In particular, the two quasiprobabilities of Refs.~\cite{solinas15,allahverdyan14} can be viewed as particular cases of a more general quasiprobability~\cite{francica22}.
To briefly introduce the problem, we recall that if the work performed on a thermally isolated quantum system is equal to the energy change of the system, then the two-projective-measurement scheme fails to describe the work statistics in the presence of initial quantum coherence in the energy basis. Basically, in this invasive scheme, where two projective measurements of the energy are performed at the initial and final times to infer the work statistics, the first measurement destroys the initial coherence in the energy basis.
A no-go theorem~\cite{perarnau-llobet17} states that there is no scheme having a probability distribution of work, which is linear with respect to the initial state, such that it reduces to the two-projective-measurement scheme for incoherent states and the average work corresponds to the average energy change. Thus, if these latter conditions are satisfied, we have to look for a quasiprobability instead of a probability. As shown in Ref.~\cite{lostaglio18}, this can be related to the contextuality of the protocol. Now a fundamental question arises: what is the quasiprobability of work which satisfies these conditions? Going in this direction, a class of quasiprobabilities has been introduced in Ref.~\cite{francica22}. However, this does not exclude the possible existence of different quasiprobabilities. Here, to give an answer we aim to deduce what the work quasiprobability is, starting from some fundamental conditions. To do this, in Sec.~\ref{sec.1} we introduce a general notion of quasiprobability in analogy to the well-known Gleason's theorem~\cite{gleason57}, starting from Ref.~\cite{bush03}. Thus, in Sec.~\ref{sec.2} we use this notion to naturally define a quasiprobability of work, and we determinate its form if some conditions need to be satisfied. To identify the quantum features and signatures of the work statistics, we discuss the negativity of the quasiprobability in relation to the initial quantum coherence in Sec.~\ref{sec. 2.5} and the contextuality of the protocol in Sec.~\ref{sec.3}. Finally, we summarize and discuss our results in Sec.~\ref{sec.c}.

\section{Quasiprobability}\label{sec.1}
In general, experimental events (or facts) are represented as effects, which are the positive operators which can occur in the range of a positive operator valued measurement, i.e., an effect is a Hermitian operator $E$ acting on the Hilbert space $\mathcal H$ such that $0\leq E \leq I$. The generalized probability measures on the set of effects are functions $E\mapsto v(E)$ with the properties
\begin{eqnarray}
\label{P1}&& 0\leq v(E) \leq 1\,,\\
\label{P2}&& v(I)=1\,,\\
\label{P3}&& v(E+F+\cdots)=v(E)+v(F)+\cdots
\end{eqnarray}
where $E+F+\cdots \leq I$. An analog of Gleason's theorem~\cite{bush03} states that, if Eqs.~\eqref{P1}-\eqref{P3} are satisfied, then the probability corresponding to the event represented by $E$ is $v(E)=\Tr{E \rho}$ for some density matrix $\rho$. Here, we try to generalize this result in the case of multiple events. For simplicity, let us focus on two events. We aim to define a function $v(E,F)$ with the properties
\begin{eqnarray}
\label{Q1}&& v(E,F)\in \R \,,\\
\label{Q2}&& v(I,E)=v(E,I)=v(E)\,,\\
\nonumber && v(E+F+\cdots,G)=v(E,G)+v(F,G)+\cdots\,,\\
\label{Q3}&& v(G,E+F+\cdots)=v(G,E)+v(G,F)+\cdots
\end{eqnarray}
where $E+F+\cdots \leq I$. In particular, we do not require that $v(E,F)$ is non-negative, so that we can get a quasiprobability, as can be expected from the Nelson theorem~\cite{nelson,breuer}.
If Eqs.~\eqref{Q1}-\eqref{Q3} are satisfied, and if $v(E,F)$ is sequentially continuous in its arguments, then the joint quasiprobability corresponding to the events represented by $E$ and $F$ is a bilinear function, in detail it is $v(E,F)=Re\Tr{E F \rho}$ for some density matrix $\rho$.

In order to prove this result, let us focus on one argument. As in Ref.~\cite{bush03}, we start to note that Eq.~\eqref{Q3} implies $v(E,F) = n v(E/n,F)$, where $n$ is a positive integer, since $v(E,F)=v(n E/n,F)=v(E/n+\cdots,F)=n v(E/n,F)$. Then, $v(p E,F) = p v(E,F)$ for any rational $p$ such that $0\leq p \leq 1$, since given $p=m/n$ we get $v(mE/n,F)=v(E/n+\cdots,F)=m v(E/n,F)=m v(E,F)/n$. To show that  $v(\alpha E,F) = \alpha v(E,F)$ for any real $\alpha$ such that $0\leq \alpha \leq 1$, it is enough to consider a sequence of rational numbers $\alpha_n\in[0,1]$ such that $\alpha_n\to \alpha$ as $n\to \infty$. We note that since $v(E,F)$ is not necessarily non-negative, we believe we cannot show it without the continuity requirement as was done in Ref.~\cite{bush03} for the case of the probability $v(E)$.
Then, by proceeding as in Ref.~\cite{bush03}, it is straightforward to show the linearity of $v$ as function of bounded Hermitian operators. Thus, $v(E,F)$ is a bilinear function, there is an operator $X_F$ such that $v(E,F)=\Tr{E X_F}$, and from Eq.~\eqref{Q2} we get $\Tr{X_F}=\Tr{F \rho}$, which implies that $X_F$ is the affine combination $X_F=q F\rho+(1-q)\rho F$.
Then, we get $v(E,F)=q \Tr{E F\rho}+(1-q)\Tr{\rho F E}$, which is real for $q=1/2$, so that $v(E,F)= Re\Tr{E F \rho}$.

For more than two events, we have a quasiprobability $v(E,F,\cdots)$ with the properties
\begin{eqnarray}
\label{Q1M}&& v(E,F,\cdots)\in \R \,,\\
\label{Q2M}&& v(I,E,F,\cdots)=v(E,I,F,\cdots)=\cdots=v(E,F,\cdots)\,,\\
\nonumber && v(E+F+\cdots,G,\cdots)=v(E,G,\cdots)+v(F,G,\cdots)+\cdots\,,\\
\label{Q3M}&& \cdots
\end{eqnarray}
where $E+F+\cdots \leq I$. Analogously, if Eqs.~\eqref{Q1M}-\eqref{Q3M} are satisfied, and if $v(E,F,\cdots)$ is sequentially continuous in its arguments, then the joint quasiprobability corresponding to the events represented by $E$, $F$, $\cdots$ is a multilinear function. In detail it is not fixed, and can be expressed as an arbitrary affine combination of $Re \Tr{X \rho}$ where $X$ are all the possible products of the effects.
In particular, for more than two events, the quasiprobability is not fixed since depends on how the events are grouped together.
For instance, let us consider the case of three events which are represented by the rank one projectors $E_i$, $E'_j$ and $E''_k$, where $E_i E_j = \delta_{i,j} E_i$ and similar equations for $E'_i$ and $E''_i$. If $E_i=E'_i$, we have only two non-trivial cases $v(E_i,E'_j,E''_k)=Re\Tr{E_iE'_jE''_k\rho}=\delta_{ij}v(E_i,E''_k)$ and $v(E_i,E''_k,E'_j)=Re\Tr{E_iE''_kE'_j\rho}$ corresponding to differently group the events $E_i$, $E'_j$ and $E''_k$.
Basically, we have decomposed the proposition $E_i\land E'_j \land E''_k$ in two different ways that are $E_i\land E'_j$, $E'_j \land E''_k$ and $E_i \land E''_k$, $E''_k\land E'_j$. Thus, in general, $Re \Tr{EFG\cdots \rho}$ will be the quasiprobability related to the joint events $E\land F$, $F\land G$, $G\land \cdots$.

It is interesting to see how a quasiprobability such defined is related to the Wigner function~\cite{wigner32}. For a one-dimensional system, if $x$ is the position and $p$ is the momentum, the Wigner function is defined as
\begin{equation}
W(x,p)=\frac{1}{\pi} \int e^{2ipy}\bra{x-y}\rho \ket{x+y}dy\,.
\end{equation}
If we consider the three events which are represented by the position projectors $\ket{x}\bra{x}$, $\ket{y}\bra{y}$ and the momentum projector $\ket{p}\bra{p}$, we get the quasiprobability
\begin{equation}
v(x,p,y)=Re \braket{x}{p}\braket{p}{y}\bra{y}\rho\ket{x}\,,
\end{equation}
which is intimately related to the Wigner function since
\begin{equation}\label{wigner ave}
\int g(x,p) W(x,p)dxdp = \int g\left(\frac{x+y}{2},p\right) v(x,p,y) dxdydp
\end{equation}
for any function $g(x,p)$.
It is worth observing that when $W(x,p)\geq 0$ for every $x$ and $p$, the quasiprobability $v(x,p,y)$ can be negative. For instance, for a pure state $\rho=\ket{\psi}\bra{\psi}$, the non-negativity of $W(x,p)$ implies that $\braket{x}{\psi}=\exp(-ax^2+bx+c)$ where $a$,$b$ and $c$ are complex variables and $Re a>0$. However, it is easy to show that by considering this state $\ket{\psi}$, $v(x,p,y)$ can be negative. In particular, by considering Eq.~\eqref{wigner ave}, we get
\begin{equation}
W(x,p) = 2 \int v(x+y,p,x-y) dy\,,
\end{equation}
so that $W(x,p)\geq 0$ if and only if $\int v(x+y,p,x-y) dy\geq 0$.

To derive a general sufficient condition such that $v(E,F,\cdots)$ is non-negative, we consider $E$, $F$, $\cdots$, as rank one projectors.
Let us start to consider the case where there are only two events represented by $E$ and $F$. If $[E,F]=0$, we get $E F = 0$ or $EF = E$, then $v(E,F)=0$ or $v(E,F)=v(E)\geq 0$.  To prove that $E F = 0$ or $EF = E$, we recall that $[E,F]=0$ implies that there is a basis of orthogonal states $\ket{i}$ such that the projectors $E$ and $F$ are diagonal. Since, $E$ and $F$ are rank one projectors, we get $E=\ket{i}\bra{i}$ and $F=\ket{j}\bra{j}$ (by labeling opportunely the basis), from which $E F = 0$ if $i\neq j$ or $E F = E$ if $i=j$.
Conversely, if $[E,\rho]=0$, we get $E \rho = p E$, where $p\geq 0$, then $v(E,F)=p \Tr{E F} \geq 0$. We can proceed analogously for more than two events, so that in general for $n$ events, $v(E,F,\cdots)$ is non-negative if there are at least $n-1$ different commutators involving the operators $\rho$, $E$, $F$, $\cdots$, that are equal to zero. This means that if $\rho$ is not incoherent with respect any projector, i.e., $[E,\rho]\neq 0$, $[F,\rho]\neq 0$, $\cdots$, then $v(E,F,\cdots)$ is non-negative if all the events are compatible, i.e., the corresponding projectors commutate each other.
On the other hand, if there are $n-2$ couples of compatible events, a negative $v(E,F,\cdots)$ implies that there is
quantum coherence in the state $\rho$ with respect to the basis of at least two events.
In general, there is quantum coherence in the state $\rho$ with respect to a basis of orthogonal states $\ket{i}$ if $\rho$ is not an incoherent state, i.e., $\rho \neq \sum_i \ket{i}\bra{i} \rho \ket{i}\bra{i}$, or equivalently if $[\rho,\ket{i}\bra{i}]\neq 0$ for some $i$.

\section{Work statistics}\label{sec.2}
We proceed by investigating the statistics of the work performed in a out-of-equilibrium process. We consider a quantum coherent process generated through a time-dependent Hamiltonian $H(t)=\sum \epsilon_k(t) \ket{\epsilon_k(t)}\bra{\epsilon_k(t)}$ where $\ket{\epsilon_k(t)}$ is the eigenstate with eigenvalue $\epsilon_k(t)$ at the time $t$. The time evolution operator is $U_{t,0}=\mathcal T e^{-i\int_0^t H(s) ds}$, where $\mathcal T$ is the time order operator and the average work $\langle w\rangle$ done on the system in the time interval $[0,\tau]$ can be identify with the average energy change
\begin{equation}\label{W1}
\langle w\rangle = \Tr{(H^{(H)}(\tau) -H(0) )\rho_0}\,,
\end{equation}
where $\rho_0$ is the initial density matrix and given an operator $A(t)$ we define the Heisenberg time evolved operator $A^{(H)}(t) = U_{t,0}^\dagger A(t) U_{t,0}$. We introduce the projectors $\Pi_i = \ket{\epsilon_i}\bra{\epsilon_i}$ and $\Pi'_k=U_{\tau,0}^\dagger \ket{\epsilon'_k}\bra{\epsilon'_k} U_{\tau,0}$, which will represent our events, where $\epsilon_i=\epsilon_i(0)$ and $\epsilon'_k=\epsilon_k(\tau)$.
Thus, we aim to find the most general quasiprobability distribution of work $p(w)$ which (W1) reproduces the two-projective-measurement scheme for an incoherent initial state, i.e., such that
\begin{equation}\label{W2}
p(w) =  \sum_{i,k} \Tr{\Pi_i \Pi'_k \rho_0} \delta(w-\epsilon'_k + \epsilon_i)
\end{equation}
if $\rho_0$ is incoherent, i.e., $\Delta(\rho_0)=\rho_0$ where $\Delta$ is the dephasing map defined as $\Delta(\rho)=\sum_i \Pi_i\rho \Pi_i$, so that $[\rho_0,\Pi_i]=0$ for every $i$. Furthermore, we require that (W2) the average work given by Eq.~\eqref{W1} can be calculated as $\langle w \rangle = \int w p(w)dw$. In particular, the nth work moment is given by $\langle w^n \rangle = \int w^n p(w)dw$.

In general, the work will be represented in terms of the events $\Pi_{i}$, $\Pi_j$, $\cdots$, $\Pi'_k$, $\cdots$, thus has a quasiprobability distribution of the form
\begin{equation}\label{eq p}
p(w) = \sum_{i,j,\cdots,k,\cdots} v(\Pi_i,\Pi_j,\cdots,\Pi'_k,\cdots) \delta(w-w(\epsilon_i,\cdots))\,,
\end{equation}
with a definite decomposition of the proposition $\Pi_i\land  \Pi_j \land \cdots$, where $w(\epsilon_i,\cdots)$ is a certain function of the eigenvalues corresponding to the events.
From the condition (W1) of Eq.~\eqref{W2}, there are only two non-trivial cases: $v(\Pi_i,\Pi'_k)=Re \Tr{\Pi'_k \Pi_i \rho_0}$ with $w(\epsilon_i,\cdots)=\epsilon'_k-\epsilon_i$, and $v(\Pi_i,\Pi'_k,\Pi_j)=Re \Tr{\Pi_i\Pi'_k \Pi_j \rho_0}$ with $w(\epsilon_i,\cdots)=\epsilon'_k-q\epsilon_i-(1-q)\epsilon_j +f_{ijk}$, where $f_{iik}=0$. In particular the first case can be obtained from the second one by considering $q=0$ or $1$ and $f_{ijk}=0$.
Concerning $f_{ijk}$, the condition (W2) that the average obtained from $p(w)$ is equal to Eq.~\eqref{W1}, gives
\begin{equation}\label{eq aus}
\sum_{i,j,k} Re \Tr{\Pi_j \Pi'_k \Pi_i \rho_0} f_{ijk}=0\,,
\end{equation}
which implies that $f_{ijk}=a_{ik}-a_{jk}+c_{ij}$ with $c_{ii}=0$. To prove it, we note that $f_{ijk}=a_{ik}+b_{jk}+c_{ij}+d_{ijk}$ where $d_{ijk}$ is a tensor. From Eq.~\eqref{eq aus}, we get
\begin{eqnarray}
\nonumber && \sum_{i,k} Re\Tr{\Pi'_k \Pi_i \rho_0} (a_{ik}+b_{ik})+ \sum_{i} \Tr{\Pi_i \rho_0} c_{ii}\\
 && + \sum_{i,j,k} Re\Tr{\Pi_j \Pi'_k \Pi_i \rho_0} d_{ijk} =0\,,
\end{eqnarray}
which needs to be satisfied for arbitrary $\Pi_i$ and $\Pi'_k$, then $a_{ik}+b_{ik}=0$, $c_{ii}=0$ and $d_{ijk}=0$. We note that if $w(\epsilon_i,\cdots)$ is a linear function of the eigenvalues, then $f_{ijk}=0$.  In particular, if it is linear then if $H(t) \to \lambda H(t)$ we get $w(\epsilon_i,\cdots) \to \lambda w(\epsilon_i,\cdots)$ and so $w \to \lambda w$. Furthermore, if $H(t) \to H(t) + e(t)$ where $e(t)$ is a c-number, we have $w\to w + e(\tau)-e(0)$, which are reasonable features of the work $w$.
Anyway, if it is non-linear there are other possible choices, for instance $a_{ik}\propto\epsilon_i^2$ and $c_{ij}=0$.
It is clear that to fix $f_{ijk}$ we have to consider a further condition. We find that $f_{ijk}=0$ if we require that (W3) the work second moment is
\begin{equation}\label{W3}
\langle w^2\rangle = \Tr{(H^{(H)}(\tau) -H(0) )^2\rho_0}\,.
\end{equation}
If $f_{ijk}=0$, the most general quasiprobability reads
\begin{equation}\label{eq pq}
p_q(w) = \sum_{k,j,i} Re\Tr{\Pi_i \Pi'_k \Pi_j \rho_0} \delta(w-\epsilon'_k + q\epsilon_i + (1-q)\epsilon_j)\,,
\end{equation}
and all these quasiprobabilities $p_q(w)$ form the class of quasiprobabilities having the form of Eq.~\eqref{eq p} and satisfying  the conditions about the reproduction of the two-projective-measurement scheme (W1), the average (W2) and the second moment (W3). In particular, an affine combination of quasiprobabilities $p_q(w)$ with different $q$ still satisfies the conditions (W1), (W2) and (W3). We note that the quasiprobability $p_q(w)$ has been introduced in Ref.~\cite{francica22}, where its relation with the initial quantum coherence has been investigated and fluctuation relations have been derived.
In contrast, if we do not require that the condition (W1) is satisfied, the number of events in Eq.~\eqref{eq p} is arbitrary, since always exists a function $w(\epsilon_i,\cdots)$ such that the condition (W2) is satisfied. For instance, we can consider the linear function $w(\epsilon_i,\cdots) = q'_1 \epsilon'_k + \cdots - q_1 \epsilon_i-\cdots$, where $q_1+q_2+\cdots = q'_1+q'_2+\cdots = 1$.

In the end, it is interesting to note that a quasiprobability having the form of Eq.~\eqref{eq p} is achieved from the condition (W2) by considering
\begin{equation}\label{eq p complex}
p(w) = \sum_{i,\cdots} \Tr{X_{i\cdots} \rho_0} \delta(w-w(\epsilon_i,\cdots))\,,
\end{equation}
where $X_{i\cdots}$ is a product of projectors, for instance $X_{ik} = \Pi'_k \Pi_i$ or $X_{iki}=\Pi_i\Pi'_k \Pi_i$. In the following we aim to prove that the condition (W2) prohibits repetitions of the projectors in the product, e.g., we cannot have $X_{iki}$.

Let us start to consider the case where only two projectors $\Pi_i$ and $\Pi'_k$ appear in the product. A product $X_{i\cdots i}$ of the form $\Pi_i \cdots \Pi_i$ does not satisfy the condition (W2) since in this case $\Tr{X_{i\cdots i} \rho_0}=\Tr{X_{i\cdots i}\Pi_i \rho_0 \Pi_i}$, then by considering an operator $\rho_0=\ket{\epsilon_i}\bra{\epsilon_j}$ with $j\neq i$, we get $\Tr{X_{i\cdots i} \rho_0}=0$ and $\int w p(w) dw= 0$, but $\langle w\rangle = \sum_k \bra{\epsilon_j}\Pi'_k \ket{\epsilon_i} \epsilon'_k$ so that the condition (W2) is not satisfied in general.
Similarly, a product $X_{k\cdots k}$ of the form $\Pi'_k \cdots \Pi'_k$ does not satisfy the condition (W2) and we can only have products of the form $\Pi_i \cdots \Pi'_k$ or $ \Pi'_k \cdots \Pi_i$.
For the case $X_{i\cdots k}=\Pi_i \cdots \Pi'_k$ we consider $\rho_0 = \Pi'_k \Pi_i$, then condition (W2) reads $\Tr{X_{i\cdots k}} w(\epsilon_i,\cdots)=\Tr{\Pi_i \Pi'_k}(\epsilon'_k-\epsilon_i)$, which implies that $X_{i\cdots k}=X_{ik}=\Pi_i \Pi'_k$. Similarly,  the only product $X_{k\cdots i}$ of the form $ \Pi'_k \cdots \Pi_i$ which satisfies the condition (W2) is $X_{ki}=\Pi'_k\Pi_i $. We proceed by considering the case where only the three projectors $\Pi_i$, $\Pi_j$ and $\Pi'_k$ appear in the product.
We cannot have products of the form $\Pi_i \cdots \Pi_i$, $\Pi_j \cdots \Pi_j$ and $\Pi'_k \cdots \Pi'_k$. We consider $X_{i\cdots k}$ of the form $\Pi_i \cdots \Pi'_k$, for $\rho_0 = \Pi'_k\Pi_i$ the condition (W2) reads $\sum_j \Tr{X_{i\cdots k}}w(\epsilon_i,\cdots) = \Tr{\Pi_i \Pi'_k}(\epsilon'_k-\epsilon_i)$ which implies that $X_{i\cdots k}=X_{ijk} = \Pi_i \Pi_j \Pi'_k$. For $X_{i\cdots j}$ of the form $\Pi_i \cdots \Pi_j$, we consider  $\rho_0=\ket{\epsilon_j}\bra{\epsilon_i}$, then the condition (W2) reads $\sum_k \bra{\epsilon_i}X_{i\cdots j}\ket{\epsilon_j} w(\epsilon_i,\cdots) = \sum_k \bra{\epsilon_i}\Pi'_k \ket{\epsilon_j} \epsilon'_k$ which implies that $X_{i\cdots j}=X_{ikj} = \Pi_i \Pi'_k \Pi_j$. Thus, in general, the condition (W2) implies that the operator $X_{i\cdots}$ is a product of the projectors without repetitions, and a quasiprobability having the form of Eq.~\eqref{eq p} is achieved by taking the real part of the quasiprobability of Eq.~\eqref{eq p complex}.

\section{Quantum coherence and negativity} \label{sec. 2.5}
The presence of quantum coherence in the initial state can lead to quantum features which are absent in a classical process, related to the quasiprobability of work.
Let us start to investigate the relation between the negativity of the quasiprobability of work $p_q(w)$ in Eq.~\eqref{eq pq} and the quantum coherence in the initial state $\rho_0$. As shown in the end of Sec.~\ref{sec.1}, since there are $n=3$ events, which are represented by the rank one projectors $\Pi_i$, $\Pi_j$ and $\Pi'_k$, if there is only one couple of compatible events, i.e., $[\Pi_i,\Pi'_k]\neq 0$ and $[\Pi_j,\Pi'_k]\neq 0$, a negative quasiprobability $v(\Pi_i,\Pi'_k,\Pi_j)<0$ implies that $\rho_0$ does not commutate with at least two projectors among $\Pi_i$, $\Pi_j$ or $\Pi'_k$. Concerning the quasiprobability of work, if $q\neq 0$ and $1$, we get that $p_{q}(w)<0$ implies $[\rho_0,\Pi_i]\neq 0$ for some $i$. Thus, in the initial state there is quantum coherence with respect to the initial energy basis.
Similarly, for the case $q=0$ or $1$, we get that $p_{0,1}(w)<0$ implies $[\rho_0,\Pi_i]\neq 0$ and $[\rho_0,\Pi'_k]\neq 0$ for some $i$ and $k$, i.e., there is initial quantum coherence with respect to the initial energy basis and final quantum coherence with respect to the final energy basis.
We note that in general the converse statements do not hold. For instance, we consider the process experimentally studied in Ref.~\cite{batalhao15}, which is a qubit with Hamiltonian $H(t) = \omega(t) (\sigma^x \cos \varphi(t)+\sigma^y \sin \varphi(t))$, where $\varphi(t) = \pi t/(2\tau)$, $\omega(t) = \omega_0 (1-t/\tau)+\omega_\tau t/\tau$ and $\sigma^x$, $\sigma^y$ and $\sigma^z$ are the Pauli matrices. We get the Heisenberg time evolved final Hamiltonian $H^{(H)}(\tau)=\omega_\tau U_{\tau,0}^\dagger \sigma^y U_{\tau,0}=\omega_\tau \hat n \cdot \vec{\sigma}$,  with $\hat n$ being a unit vector. For studying the effect of the initial quantum coherence we take the initial density matrix $\rho_0 = I/2+ r \sigma^x/2 + r' \hat n \cdot \vec{\sigma}/2$, with $r$ and $r'$ real parameters such that $(r+n_x r')^2+{r'}^2(n_y^2+n_z^2)\leq 1$. We note that $[\rho_0,\Pi_i]\neq 0$ and $[\rho_0,\Pi'_k]\neq 0$ if $r\neq 0$, $r'\neq 0$ and $\hat n \cdot \vec \sigma \neq \sigma_x$. In general, we get the quasiprobabilities
\begin{equation}
v(\Pi_i,\Pi'_k) = \frac{1}{2}(1+s_i r + s'_k r')\Tr{\Pi_i \Pi'_k}\,,
\end{equation}
where we have defined the signs $s_i=\pm$ and $s'_k = \pm$ such that $\sigma^x \Pi_i = s_i \Pi_i$ and $\hat n \cdot \vec{\sigma}\Pi'_k = s'_k \Pi'_k$. Thus, $v(\Pi_i,\Pi'_k)$ can be negative if $1+s_i r + s'_k r'<0$, but there are also parameters $r\neq 0$ and $r'\neq 0$ small enough such that $1+s_i r + s'_k r'\geq 0$, so that $v(\Pi_i,\Pi'_k)\geq 0$ and $p_{0,1}(w)\geq 0$.
In contrast, for $q\neq 0$ and $1$, the quasiprobability of work is defined in terms of the quasiprobabilities $v(\Pi_i,\Pi'_k,\Pi_j)=Re \Tr{\Pi_i\Pi'_k \Pi_j \rho_0}$, which are negative for some $i$ and $j\neq i$ if $[\rho_0,\Pi_i]\neq 0$ for some $i$. To prove it, it is enough to consider that $v(\Pi_i,\Pi'_k,\Pi_j)=\delta_{i,j}\Tr{\Pi_i \Pi'_k \Delta(\rho_0)} + \delta v_{ikj}$, where $\delta v_{ikj}=Re \Tr{\Pi_i \Pi'_k \Pi_j (\rho_0-\Delta(\rho_0)}$ needs to be negative for some $i$ and $j\neq i$ because $\sum_{i,j,k} \delta v_{ikj}=0$, since $\sum_{i,j,k} v(\Pi_i,\Pi'_k,\Pi_j) = 1$ and $\sum_{i,k} \Tr{\Pi_i \Pi'_k \Delta(\rho_0)}=1$.
However,  this does not rule out that we can get both $[\rho_0,\Pi_i]\neq 0$ and $p_{q}(w)\geq 0$ for some $q\neq 0$ and $1$, as we will discuss in the next section, where an explicit example for $q=1/2$ is also given.

\section{Contextuality} \label{sec.3}
Having determined the form of the quasiprobability of work and characterized the negativity in relation to the initial quantum coherence, now we can focus on the problem if there is a non-contextual hidden variables model which satisfies the conditions about the reproduction of the two-projective-measurement scheme (W1), the average (W2) and the second moment (W3).
To introduce the concept of contextuality at an operational level (see, e.g., Ref.~\cite{lostaglio18}), we consider a set of preparations procedures $P$ and measurements procedures $M$ with outcomes $k$, so that we will observe $k$ with probability $p(k|P,M)$. We aim to reproduce the statistics by using a set of states $\lambda$ that are random distributed in the set $\Lambda$ with probability $p(\lambda|P)$ every time the preparation $P$ is performed. If, for a given $\lambda$, we get the outcome $k$ with the probability $p(k|\lambda,M)$, we are able to reproduce the statistics if
\begin{equation}\label{eq con}
p(k|P,M) = \int_\Lambda p(\lambda|P) p(k|\lambda,M)d\lambda\,,
\end{equation}
and the protocol is called universally non-contextual if $p(\lambda|P)$ is a function of the quantum state alone, i.e., $p(\lambda|P)=p(\lambda|\rho_0)$, and $p(k|\lambda,M)$ depends only on the POVM element $M_k$ associated to the corresponding outcome of the measurement $M$, i.e., $p(k|\lambda,M)=p(k|\lambda,M_k)$. In our case, the outcome $k$ corresponds to the work $w_k$, and  if the protocol is non-contextual the work distribution can be expressed as
\begin{equation}\label{eq non cont}
p(w) = \sum_k p(k|P,M) \delta(w-w_k)\,,
\end{equation}
where $p(k|P,M)$ is given by Eq.~\eqref{eq con} with $p(\lambda|P)=p(\lambda|\rho_0)$ and $p(k|\lambda,M)=p(k|\lambda,M_k)$, so that for a negative quasiprobability of work we cannot have a non-contextual protocol.
Thus, a process that cannot be reproduced within any non-contextual protocol will exhibit genuinely non-classical features.
We point out that the negativity of $v(\Pi_i,\Pi'_k)$ or $v(\Pi_i,\Pi'_k,\Pi_j)$ alone is not sufficient to guarantee contextuality, i.e., that there is no non-contextual protocol that satisfies the conditions (W1), (W2) and (W3), and in principle we have to verify  that there is no some quasiprobability of work which is non-negative and so non-contextual.
To prove this we give a counterexample. We consider a one dimensional system, with initial Hamiltonian $H(0)=\alpha x$ and final Hamiltonian $H(\tau)=\beta p^2$. In this case we consider the projectors $\Pi_x=\ket{x}\bra{x}$ and $\Pi'_p=\ket{p}\bra{p}$. It is easy to show that for $q=1/2$ the work quasiprobability of Eq.~\eqref{eq pq} can be expressed in terms of the Wigner function as
\begin{equation}\label{eq 1/2}
p_{1/2}(w) = \int dx dp W(x,p) \delta(w-\beta p^2+\alpha x)\,.
\end{equation}
If we consider the initial wave function $\braket{x}{\psi}=\exp(-ax^2+bx+c)$, we get $W(x,p)\geq 0$, thus the protocol is non-contextual.
In detail, we have $\lambda=(x,p)$, $p(\lambda|\rho_0)=W(x,p)$, $p(k|\lambda,M_k)=\delta(w_k-\beta p^2 +\alpha x)$ and a continuous variable $w_k$, so that from Eq.~\eqref{eq non cont} we get Eq.~\eqref{eq 1/2}.
However, for this state it is easy to verify that the quasiprobabilities $v(\Pi_x,\Pi'_p)$ and $v(\Pi_x,\Pi'_p,\Pi_y)$ can take negative values, so we cannot say in general if a protocol corresponding to an arbitrary $q$ exhibits contextuality by only looking on the negativity of these quasiprobabilities.
In particular, one would expect the negativity of $v(\Pi_i,\Pi'_k)$ to give a sufficient condition for contextuality for any $q$, since $v(\Pi_i,\Pi'_k)<0$ implies that there is initial quantum coherence with respect to the initial energy basis, so that $v(\Pi_i,\Pi'_k,\Pi_j)<0$ for some $i$ and $j\neq i$. However, some negative $v(\Pi_i,\Pi'_k,\Pi_j)<0$ do not guarantee the negativity of $p_q(w)$ for any $q$.
We note that this situation is in contrast with the one examined in Ref.~\cite{lostaglio18}, where only the case $q=0$ and $1$ has been considered and the negativity of $v(\Pi_i,\Pi'_k)$ implies that the corresponding protocol is contextual.
Basically, this is due to the fact that, although can exist a unique couple of indexes $i$ and $k$ such that $\epsilon'_k-\epsilon_i=w$, for $q\neq0$ and $1$ we can have different triples of indexes $i$, $j$ and $k$ such that $\epsilon'_k-q\epsilon_i-(1-q)\epsilon_j=w$, so that the quasiprobability associated to this value $w$ is obtained by summing $v(\Pi_i,\Pi'_k,\Pi_j)$ over these triples, and of course the negativity of some quasiprobabilities $v(\Pi_i,\Pi'_k,\Pi_j)$ does not guarantee the negativity of the sum.

\section{Conclusions}\label{sec.c}
Understanding how to describe nature from a statistical point of view is a problem of fundamental importance. For instance, by characterizing the work statistics, it is possible to derive some fluctuation theorems related to the second law of thermodynamics (see, e.g., Refs.~\cite{jarzynski97,crooks99}), which clarifies the importance of this study. In quantum systems, by focusing on the work, this description is less obvious than the classical case and it is still a matter of discussion (see, e.g., Ref.~\cite{book19}).
With the aim of helping to clarify what the work statistics can be, we identified the form of the work quasiprobability which is compatible with some fundamental conditions.
In particular, we introduced a general notion of quasiprobability by giving an argumentation analogous to the Gleason's theorem.
Recently, the energy-change statistics in a generic open quantum system has been characterized by means of the Kirkwood-Dirac quasiprobability~\cite{lostaglio22}. By following this approach, the condition that the quasiprobability is real can be relaxed, however in our approach this condition is important since ensures a unique quasiprobability related to joint events.
We note that these quasiprobabilities are experimentally accessible (e.g., see Appendix~\ref{appendix}), for instance, recently, the Margenau-Hill distribution of the work (corresponding to $q=0$ or $1$) has been experimentally measured in a driven three-level quantum system~\cite{hernandez-gomez22}.
Beyond their foundational implications, we believe our findings can be useful to characterize the quantum signatures in the work statistics. In particular, due to the initial quantum coherence in the energy basis we can get (not necessarily) a contextual protocol. However, we proved that for a given initial state, it is not said that all work quasiprobabilities exhibit contextuality.

\subsection*{Acknowledgements}
The author acknowledges financial support from the project BIRD 2021 "Correlations, dynamics and topology in long-range quantum systems" of the Department of Physics and Astronomy, University of Padova.

\appendix

\section{Measuring the quasiprobability}\label{appendix}
The quasiprobability of work can be experimentally measured by using a full-counting statistics approach as shown in Ref.~\cite{francica22}.  Here, we observe that the quasiprobabilities $v(\Pi_i,\Pi'_k,\Pi_j)=Re \Tr{\Pi_i \Pi'_k \Pi_j \rho_0}$ can be measured by exploiting a detector $D$ with an Hilbert space having at least the same dimension of the Hilbert space of the system. The detector is prepared in the initial state $\ket{0}$, so that the initial state of the system and the detector is
\begin{equation}
\rho_0 \otimes \ket{0}\bra{0} = \sum_{i,j} \Pi_i \rho_0 \Pi_j \otimes \ket{0}\bra{0}\,.
\end{equation}
First, we perform a unitary operation $U_I$ on the total system, defined such that $U_I \ket{\epsilon_i}\otimes\ket{0} = \ket{\epsilon_i}\otimes \ket{\phi_i}$, where $\ket{\phi_i}$ are orthogonal states of the detector. Then, we realize the out-of-equilibrium process, i.e., we perform the unitary $U_{\tau,0}\otimes I$ on the total system. In the end, we perform a measurement of the final energy of the system. The detector collapses into the state
\begin{equation}
\rho_{D|k} = \frac{1}{p_k}\sum_{i,j} \Tr{\Pi'_k \Pi_i \rho_0 \Pi_j}  \ket{\phi_i}\bra{\phi_j}
\end{equation}
with probability $p_k = \sum_i \Tr{\Pi'_k \Pi_i \rho_0 \Pi_i}$. Thus, by performing measurements on the detector states $\rho_{D|k}$ it is possible to determinate the matrix elements $\Tr{\Pi'_k \Pi_i \rho_0 \Pi_j}/p_k$, from which we get the quasiprobabilities $v(\Pi_i,\Pi'_k,\Pi_j)$.


\begin{thebibliography}{99}
\bibitem{talkner07} P. Talkner, E. Lutz and P. H\"{a}nggi, Phys. Rev. E 75, 050102(R) (2007).
\bibitem{deffner16} S. Deffner, J. P. Paz and W. H. Zurek Phys. Rev. E 94, 010103(R) (2016).
\bibitem{talkner16} P. Talkner, and P. H\"{a}nggi, Phys. Rev. E 93, 022131 (2016).
\bibitem{allahverdyan14} A. E. Allahverdyan, Phys. Rev. E 90, 032137 (2014).
\bibitem{solinas15} P. Solinas, and S. Gasparinetti, Phys. Rev. E 92, 042150 (2015).
\bibitem{miller17} H. J. D. Miller, and J. Anders, 2017 New J. Phys. 19 062001.
\bibitem{francica22} G. Francica, Phys. Rev. E 105, 014101 (2022).
\bibitem{perarnau-llobet17} M. Perarnau-Llobet, E. B\"{a}umer, K. V. Hovhannisyan, M. Huber and A. Acin, Phys. Rev. Lett. 118, 070601 (2017).
\bibitem{lostaglio18} M. Lostaglio, Phys. Rev. Lett. 120, 040602 (2018).
\bibitem{gleason57} A.M. Gleason, J. Math. Mech. 6, 885 (1957).
\bibitem{bush03} P. Busch, Phys. Rev. Lett. 91, 120403 (2003).
\bibitem{nelson} E. Nelson, {\it Dynamical Theories of Brownian Motion}, Princeton University Press, Princeton (1967).
\bibitem{breuer} H.-P. Breuer, F. Petruccione, {\it The Theory Of Open Quantum Systems}, Oxford University Press (2002).
\bibitem{wigner32} E. Wigner, Phys. Rev. 40, 749 (1932).
\bibitem{batalhao15} T. B. Batalh\~{a}o, A. M. Souza, R. S. Sarthour, I. S. Oliveira, M. Paternostro, E. Lutz and R. M. Serra, Phys. Rev. Lett. 115, 190601 (2015).
\bibitem{jarzynski97} C. Jarzynski, Phys. Rev. Lett. 78, 2690 (1997).
\bibitem{crooks99} G. Crooks, Phys. Rev. E, 60, 2721 (1999).
\bibitem{book19} {\it Thermodynamics in the Quantum Regime}, edited by F. Binder, L. A. Correa, C. Gogolin, J. Anders, and G. Adesso (Springer, Cham, 2019).
\bibitem{lostaglio22} M. Lostaglio, A. Belenchia, A. Levy, S. Hern\'{a}ndez-G\'{o}mez, N. Fabbri, S. Gherardini, arXiv:2206.11783 (2022).
\bibitem{hernandez-gomez22} S. Hern\'{a}ndez-G\'{o}mez, S. Gherardini, A. Belenchia, M. Lostaglio, A. Levy, N. Fabbri, arXiv:2207.12960 (2022).
\end{thebibliography}
\end{document}